\documentclass[aps,pra,showpacs,twocolumn]{revtex4-1}
\usepackage{amssymb}
\usepackage{amsmath}
\usepackage{graphicx}
\usepackage{epsfig}

\begin{document}

\title{Momentum-independent reflectionless transmission in the non-Hermitian
time-reversal symmetric system}
\author{X. Z. Zhang and Z. Song}
\email{songtc@nankai.edu.cn}
\affiliation{School of Physics, Nankai University, Tianjin 300071, China}

\begin{abstract}
We theoretically study the non-Hermitian systems, the non-Hermiticity of
which arises from the unequal hopping amplitude (UHA) dimers. The
distinguishing features of these models are that they have full real spectra
if all of the eigenvectors are time-reversal ($\mathcal{T}$) symmetric
rather than parity-time-reversal ($\mathcal{PT}$) symmetric, and that their
Hermitian counterparts are shown to be an experimentally accessible system,
which have the same topological structures as that of the original ones but
modulated hopping amplitudes within the unbroken region. Under the
reflectionless transmission condition, the scattering behavior of
momentum-independent reflectionless transmission (RT) can be achieved in the
concerned non-Hermitian system. This peculiar feature indicates that, for a
certain class of non-Hermitian systems with a balanced combination of the RT
dimers, the defects can appear fully invisible to an outside observer.
\end{abstract}

\maketitle


\section{Introduction}

\label{sec_intro}One of the fundamental characteristics of quantum mechanics
is associated with the Hermiticity of physical observables. In the case of
the Hamiltonian operator, this requirement not only implies real eigen
energies but also guarantees probability conservation \cite{Shankar}.
However, a decade ago it was observed that many non-Hermitian Hamiltonians
possess real spectra \cite{Bender} and a pseudo-Hermitian Hamiltonian
connects with its equivalent Hermitian Hamiltonian via a similarity
transformation \cite{Ali02,Ali04}. Thus quantum theory based on
non-Hermitian Hamiltonian was established \cite%
{Ali04,Ahmed,Berry,Heiss,Bender02,Bender07,Jones,Muga,Dorey,Ali10}. Most
recently, the growth of interest in the theory of parity-time ($\mathcal{PT}$%
) symmetric potentials was originated by suggestions of implementation of $%
\mathcal{PT}$ symmetry in a waveguide with gain and absorption \cite%
{Ruschhaupt}, which was based on the analogy between quantum mechanics and
paraxial optics where the refractive index plays the role of the potential
in the Schr\"{o}dinger equation \cite{R. El-Ganainy,K. G.
Makris,Christodoulides,Z. H. Musslimani,S. Klaiman,S. Longhi,Guo,H.
Schomerus,LonghiLaser,YDChong,Keya}. By exploiting optical modulation of the
refractive index in the complex dielectric permittivity plane and
engineering both optical absorption and amplification, parity-time
metamaterials can lead to a series of intriguing optical phenomena and
devices, such as absorption enhanced transmission \cite{Guo}, double
refraction, dynamic power oscillations and nonreciprocity of light
propagation \cite{K. G. Makris,Z. H. Musslimani,S. Klaiman,S.
Longhi,Zheng,Graefe,Miroshnichenko}, which can be used to realize a new
generation of on-chip isolators and circulators \cite{Hamidreza}. Other
intriguing results within the framework of $\mathcal{PT}$ optics include the
study of Bloch oscillations \cite{S. Longhi}, coherent perfect
absorber-lasers \cite{LonghiLaser,Ge,Li}, and nonlinear switching structures
\cite{Sukhorukov}.

Over the last few years, $\mathcal{PT}$-symmetric lattice models have become
a focal point of research \cite{MZnojil,Bendix,SLonghi,ZLi,Bousmina,Liang
Jin,Joglekar,Joglekar1,Joglekar2}. Lattice models, in general, are popular
in physics due to their versatility, availability of exact solutions \cite%
{Onsager}, the absence of an ultraviolet divergence \cite{Kogut}, the
ability to capture counterintuitive phenomena that have no counterparts in
the continuum theories \cite{Winkler}, as well as the experimental
accessibility. A general non-Hermitian tight-binding network is constructed
topologically by the sites and the various non-Hermitian connections between
them. There are three types of basic non-Hermitian clusters leading to the
non-Hermiticity of a discrete non-hermitian\ system: i) complex on-site
potential denoted as $\left\vert V\right\vert e^{i\varphi }a_{j}^{\dagger
}a_{j}$\ with $a_{j}$\ being boson (fermion) operator; ii) non-hermitian
dimer denoted as $\left\vert J\right\vert e^{i\varphi }$($a_{j}^{\dagger
}a_{j+1}+a_{j+1}^{\dagger }a_{j}$); iii) unequal hopping amplitude (UHA)
dimer denoted as $\mu a_{j}^{\dagger }a_{j+1}+\nu a_{j+1}^{\dagger }a_{j}$
with $\mu \neq \nu $ being real numbers. When $\mu =1/\nu $, UHA dimer
reduces to asymmetric hopping amplitude (AHA) dimer denoted as $%
e^{h}a_{j}^{\dagger }a_{j+1}+e^{-h}a_{j+1}^{\dagger }a_{j}$ with asymmetric
parameter $h$ being real number, which is used in modeling a delocalization
phenomenon relevant for the vortex pinning in superconductors \cite{Hatano}.
The former two types of non-Hermitian clusters violate $\mathcal{T}$
symmetry, while the last one does not. The systems containing the former two
types of clusters have been mainly focused on lattices possessing $\mathcal{%
PT}$ symmetry and were framed in the context of non-Hermitian quantum
mechanics \cite{Bendix,CMBender,SLonghi,Liang
Jin,Joglekar,Joglekar1,Joglekar2,ZXZ1,ZXZ3}. However, much less attention
has been paid to the property of the non-Hermitian system with the third
type of the cluster.

It is the aim of this paper to study the characteristic of the non-Hermitian
systems with UHA dimers, which possess the same properties with a
pseudo-Hermitian Hamiltonian. We show that the system has full real spectrum
if all of the eigenvectors have to respect $\mathcal{T}$-rather than $%
\mathcal{PT}$-reversal symmetry, and its Hermitian counterpart exhibits the
same topological structure as that of the original one within the unbroken
region. We also find out the behavior of the $k$-independent reflectionless
transmission (RT) for the concerned non-Hermitian scattering center under
the RT condition. This is made possible by considering the invisibility of a
state in the system with a balanced combination of the reflectionless
transmission dimers.

This paper is organized as follows. Section \ref{sec_scatter}, presents the
exact analytical solution of the scattering problem for the concerned
non-Hermitian scattering center. Section \ref{sec_connection} consists of
two exactly solvable examples to illustrate the connection with the
pseudo-Hermiticity of the $\mathcal{PT}$-symmetric non-Hermitian system.
Section \ref{sec_dynamic} is devoted to the numerical simulation of the wave
packet dynamics to demonstrate the phenomena of invisibility in two
dimensional case. We conclude the paper with a brief discussion in Section %
\ref{sec_summary}.

\section{Reflectionless transmission condition}

\label{sec_scatter}
\begin{figure}[tbp]
\includegraphics[ bb=24 43 534 800, width=0.475\textwidth, clip]{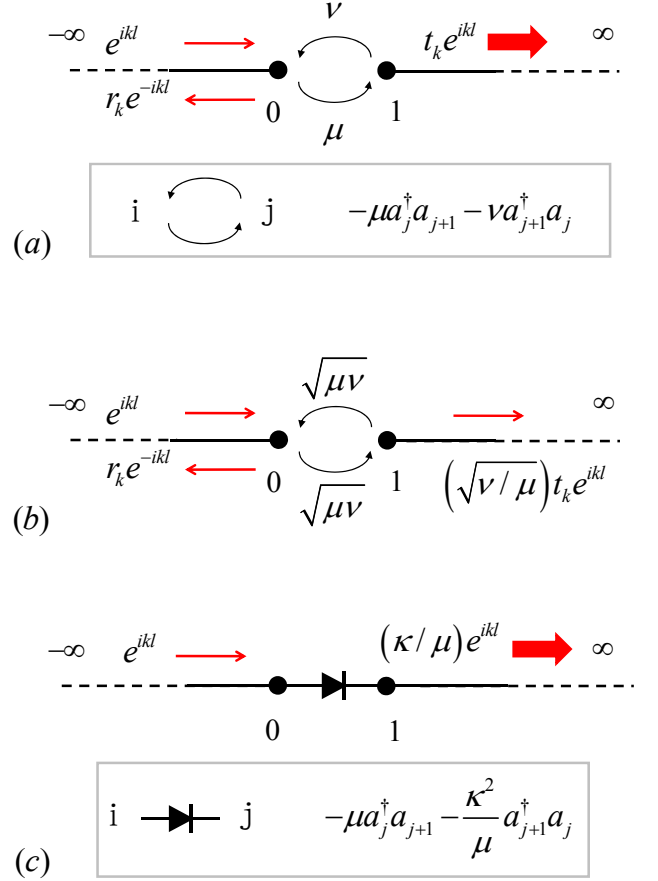}
\caption{(Color online) Schematic illustration of the configurations of the
concerned non-Hermitian networks. (a) An UHA dimer with hopping constant $\left\{ \protect\mu ,\protect\nu \right\} $ as a scattering center is
embedded in an infinite chain with hopping constant $\protect\kappa $. It is
shown that the scattering matrix of such a non-Hermitian scattering center
is similar to that of a Hermitian dimer with hopping constant $\protect\sqrt{\protect\mu \protect\nu }$, which is illustrated in (b). They share the same
reflection amplitudes and isomorphic transmission amplitudes. (c) In the
case of $\protect\kappa =\protect\sqrt{\protect\mu \protect\nu }$ ($\protect\mu <\protect\nu $), the scattering center acts as a reflectionless
amplifier, which is denoted by the amplifier symbol \textbf{\textquotedblleft }$\blacktriangleright \mathbf{\shortmid }$\textquotedblright }
\label{fig1}
\end{figure}

In order to understand the differences between the
characters of an UHA dimer with other types of non-Hermitian clusters, it is
appropriate to begin with the scattering problem for incident plane wave. As
is normally done in previous works \cite{JL1,JL2}, one can embed the
scattering center in an infinite lead, which is illustrated schematically in
Fig. \ref{fig1} (a). The Hamiltonian of the concerned scattering
tight-binding network has the form%
\begin{equation}
H_{\mathrm{Scatt}}=H_{\mathrm{L}}+H_{\mathrm{C}}+H_{\mathrm{R}}\text{,}
\end{equation}%
where%
\begin{eqnarray}
H_{\mathrm{L}} &=&-\kappa \sum_{j=-\infty }^{-1}\left( a_{j}^{\dagger
}a_{j+1}+\text{\textrm{H.c.}}\right) , \\
H_{\mathrm{R}} &=&-\kappa \sum_{j=1}^{\infty }\left( a_{j}^{\dagger }a_{j+1}+%
\text{\textrm{H.c.}}\right) ,
\end{eqnarray}%
represents the left ($H_{\mathrm{L}}$) and right ($H_{\mathrm{R}}$)
waveguides with real $\kappa $ and%
\begin{equation}
H_{\mathrm{C}}=-\left( \mu a_{0}^{\dagger }a_{1}+\nu a_{1}^{\dagger
}a_{0}\right)  \label{Hc}
\end{equation}%
describes an UHA dimer as a scattering center when $\mu \neq \nu $. Here $%
a_{j}^{\dagger }$ and $a_{j}$ denote the boson or fermion creation and
annihilation operators on the $j$th site, respectively, satisfying the
canonical commutation relations $\left[ a_{i},a_{j}^{\dagger }\right] _{\pm
}=\delta _{ij}$ and $\left[ a_{i},a_{j}\right] _{\pm }=0$. Here, for the
sake of simplicity, we only concern the case with positive $\mu $ and $\nu $%
. We first deal with the single-particle case and we will see that the
generalization to the case of an arbitrary number of particles is
straightforward.

In the single-particle subspace, for an incident plane wave with momentum $k$
incoming from the left direction with energy $E=-2\kappa \cos k$, the
scattering wave function $\left\vert \psi _{k}\right\rangle $ can be
obtained by the Bethe ansatz method. The wave function has the form%
\begin{equation}
\left\vert \psi _{k}\right\rangle =\sum_{j=-\infty }^{\infty }f^{k}\left(
j\right) a_{j}^{\dagger }\left\vert \text{\textrm{Vac}}\right\rangle
\end{equation}%
where the scattering wave function $f^{k}\left( j\right) $ is in the form of
\begin{equation}
f^{k}\left( j\right) =\left\{
\begin{array}{c}
e^{ikj}+r_{k}e^{-ikj}\text{ }\left( j\leqslant 0\right) \\
t_{k}e^{ikj}\text{ }\left( j\geqslant 1\right)%
\end{array}%
\right. .
\end{equation}%
Here $r_{k}$, $t_{k}$ are the reflection and transmission amplitudes of the
incident wave with momentum $k$, which are what we are concerned with most
in this paper. By substituting the wave function $\left\vert \psi
_{k}\right\rangle $ into the Schr\"{o}dinger equation,%
\begin{eqnarray}
-\kappa f^{k}\left( j-1\right) -\kappa f^{k}\left( j+1\right) =Ef^{k}\left(
j\right) , &&  \notag \\
j\notin \left( -1,2\right) \text{,} &&  \notag \\
-\kappa f^{k}\left( -1\right) -\mu f^{k}\left( 1\right) =Ef^{k}\left(
0\right) \text{,} && \\
-\kappa f^{k}\left( 2\right) -\nu f^{k}\left( 0\right) =Ef^{k}\left(
1\right) \text{,} &&  \notag
\end{eqnarray}%
which lead to%
\begin{eqnarray}
r_{k} &=&\frac{\kappa ^{2}-\mu \nu }{\mu \nu -\kappa ^{2}e^{-i2k}}\text{,}
\label{r_k} \\
t_{k} &=&\kappa \frac{\nu \left( 1-e^{-i2k}\right) }{\mu \nu -\kappa
^{2}e^{-i2k}}\text{.}  \label{t_k}
\end{eqnarray}%
According to traditional quantum mechanics, a scattering matrix (S-matrix),
which relates to the initial state and the final state of a physical system
undergoing a scattering process, reveals almost complete features of the
scattering center. Then the corresponding S-matrix can be written as

\begin{equation}
\widetilde{S}_{k}=\left(
\begin{array}{cc}
r_{k} & \frac{\mu }{\nu }t_{k} \\
t_{k} & r_{k}e^{-2ik}%
\end{array}%
\right) ,
\end{equation}%
where the tilt represents the dimer being non-Hermitian. To characterize the
feature of $\widetilde{S}_{k}$, we consider the S-matrix at the Hermitian
point with $\mu =\nu =\sqrt{\mu \nu }$, which is unitary and has the form

\begin{equation}
S_{k}=\left(
\begin{array}{cc}
r_{k} & \sqrt{\mu /\nu }t_{k} \\
\sqrt{\mu /\nu }t_{k} & r_{k}e^{-2ik}%
\end{array}%
\right) .
\end{equation}%
We notice that although $\widetilde{S}_{k}$ is non-Hermitian and not
unitary, the reflection amplitudes are the same and the transmission
coefficients are proportional to that of $S_{k}$. These features show that
the UHA dimer $\left\{ \mu ,\nu \right\} $ is familiar to a Hermitian dimer
with hopping amplitude $\sqrt{\mu \nu }$. Furthermore, in the case of $%
\kappa ^{2}=\mu \nu $, we have $r_{k}=0$\ and $t_{k}=\kappa /\mu $, which is
termed as the reflectionless transmission condition. It shows that an
incident plane wave can completely pass through the dimer and the amplitude
of the transmitted wave is amplified by a factor of $\kappa /\mu $.
Inversely, an incident wave from right is\text{ diminished by a factor of }$%
\mu /\kappa $. This procedure is illustrated schematically in Fig. \ref{fig1}
(c). Before further discussion of application of the obtained result, two
distinguishing features need to be mentioned. Firstly, the $H_{\mathrm{C}}$
is a non-Hermitian system can have full real spectrum. Secondly, the feature
of RT is $k$-independent.

\begin{figure}[tbp]
\includegraphics[ bb=65 298 564 650, width=0.475\textwidth, clip]{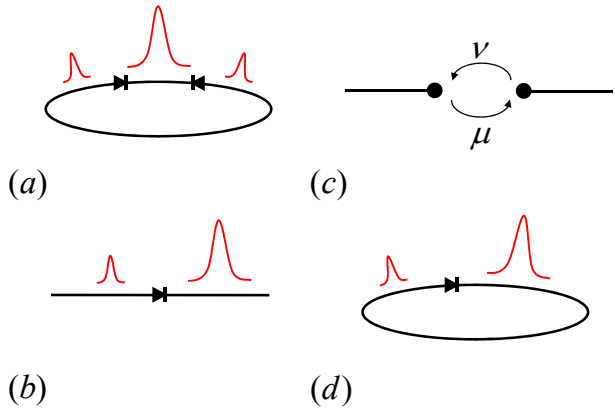}
\caption{(Color online) Schematic illustration for the exemplified
non-Hermitian systems containing UHA dimer. The reality of the spectrum can
be demonstrated from the dynamics of a wave packet in systems. (a) A ring
system containing two RT dimers. When the two dimers are balance, an
arbitrary wave packet can move in the ring persistently, which indicates the
steady current and real spectrum.\ (b) and (c) are chain systems containing
single UHA dimer.} \label{fig2}
\end{figure}


Now we investigate the S-matrix of multi-RT-dimer case. Consider an
scattering center consisting of $n$ RT dimers with hopping amplitude $%
\left\{ \mu _{i},\kappa ^{2}/\mu _{i}\right\} $, $i\in \left[ 1,n\right] $.
Then the S-matrix of $i$th dimer is%
\begin{equation}
M_{i}=\left(
\begin{array}{cc}
0 & \mu _{i}/\kappa \\
\kappa /\mu _{i} & 0%
\end{array}%
\right) ,
\end{equation}%
where we use $M$ to denote the matrix under the RT condition. One can go
further, by direct analysis, to yield the S-matrix of the whole center%
\begin{equation}
M^{\left( n\right) }=\left(
\begin{array}{cc}
0 & \prod\limits_{i=1}^{n}\left( \mu _{i}/\kappa \right) \\
\prod\limits_{i=1}^{n}\left( \kappa /\mu _{i}\right) & 0%
\end{array}%
\right) .
\end{equation}

Remarkably, in the case of unitary transmission, i.e.,%
\begin{equation}
\prod\limits_{i=1}^{n}\mu _{i}=\kappa ^{n},  \label{k_n}
\end{equation}%
we have%
\begin{equation}
M^{\left( n\right) }\left( M^{\left( n\right) }\right) ^{\dag }=1
\label{S unitary}
\end{equation}%
which is the characteristic of a Hermitian scattering center. It indicates
that a multi-dimer ($n>1$) can be equivalent to a Hermitian system. The
physics of such a phenomenon is clear: It is the result of the balance
between amplification and diminishment dimers. Nevertheless, unlike the
system with imaginary potentials,\ such a balance does not require the
symmetry of the dimer. This point can be seen from the condition in Eq. (\ref%
{k_n}), which indicates that the balance obey the commutative law\textbf{.}

In the following, we will further extend the discussion in two aspects.
Firstly, the obtained result from the single-particle scattering problem can
be extended to many-particle case and exploited to construct the finite-size
non-Hermitian system with fully real spectrum, which is beyond the framework
of the $\mathcal{PT}$-symmetric quantum mechanics. Secondly, the $k$%
-independent Hermitian dynamical behavior of the RT non-Hermitian scattering
center can be applicable to the realization of invisibility in higher
dimensional lattice.


\begin{figure}[tbp]
\includegraphics[ bb=66 186 531 760, width=0.475\textwidth, clip]{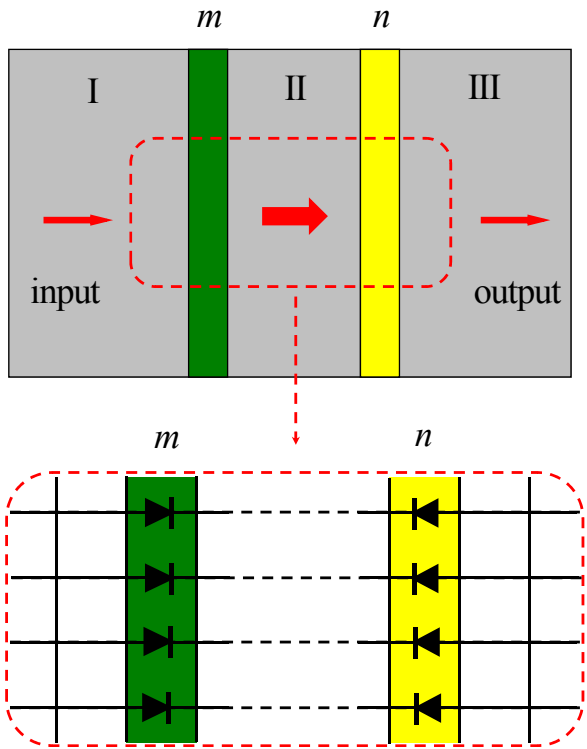}
\caption{(Color online) Schematic illustration of a two-dimensional square
lattice with $N$ sites along the horizontal direction and $M$ sites along
the vertical direction It contains two non-Hermitian layers at $m$th and $n$th columns ($1<m<n<N$), which separate the square into three regions. The
two layers act a reflectionless amplifiers. One should note that the symbol
of amplifiers are in opposite direction.} \label{fig3}
\end{figure}


\section{Non-Hermitian systems with real spectra}

\label{sec_connection}Now we investigate the connection between the UHA
dimer and the pseudo-Hermiticity of the $\mathcal{PT}$-symmetric
non-Hermitian system. Consider a finite non-Hermitian system which is
constructed by an arbitrary Hermitian lattice with time-reversal symmetry
(without magnetic flux threaded in) embedded by the UHA dimers. It becomes a
non-Hermitian system with time-reversal symmetry, i.e.,

\begin{eqnarray}
\mathcal{T}H\mathcal{T}^{-1} &=&H^{\ast }=H \\
H^{\dag } &\neq &H  \notag
\end{eqnarray}%
It has been shown \cite{AM} that if a Hamiltonian $H$ has a symmetry given
by an anti-linear operator such as time-reversal operator $\mathcal{T}$,
then either the eigenvalues of $H$ are real or they come in complex
conjugate pairs. Furthermore, an eigenvalue of $H$ is real that denotes the
corresponding eigenvector is invariant under the action of $\mathcal{T}$. It
is presumable that the analysis for a $\mathcal{PT}$-symmetric non-Hermitian
system can be applied to the one with only $\mathcal{T}$ symmetry. In the
following we will demonstrate this point by solving concrete examples
explicitly. We will focus the investigation on three characteristic
features: i) critical behavior near the exceptional point; ii) symmetry
breaking; iii) biorthogonal bases.

It is natural to connect the Hermiticity of a non-Hermitian scattering
center to the reality of the spectrum of a finite non-Hermitian system which
contains the center as building blocks. Before we start with our
investigation, we first introduce the method of operator transformation.
Consider a generalized Hamiltonian, which is written as,%
\begin{equation}
H_{\mathrm{Gen}}=H_{\mathrm{Scatt}}+\sum_{j=-\infty }^{\infty
}V_{j}a_{j}^{\dagger }a_{j}\text{,}
\end{equation}%
where $V_{j}$\ is an arbitrary real on-site potential. Inspired by the
solution of single-particle in Eqs. (\ref{r_k}) and (\ref{t_k}), one can
rewrite the Hamiltonian $H_{\mathrm{Gen}}$ as

\begin{eqnarray}
h_{\mathrm{Gen}} &=&-\kappa \sum_{j=-\infty }^{\infty }\left( \overline{b}%
_{j}b_{j+1}+\overline{b}_{j+1}b_{j}+V_{j}\overline{b}_{j}b_{j}\right)
\label{h_Gen} \\
&&+\left( \kappa -\sqrt{\mu \nu }\right) \left( \overline{b}_{0}b_{1}+%
\overline{b}_{1}b_{0}\right) ,  \notag
\end{eqnarray}%
by taking the transformation%
\begin{equation}
b_{j}\text{ }\left( \overline{b}_{j}\right) =\left\{
\begin{array}{cc}
\sqrt{\frac{\mu }{\nu }}a_{j}\text{ }\left( \sqrt{\frac{\nu }{\mu }}%
a_{j}^{\dagger }\right) & \left( j>0\right) \\
a_{j}\text{ }\left( a_{j}^{\dagger }\right) & \left( j\leqslant 0\right)%
\end{array}%
\right. .  \label{bj}
\end{equation}%
Associating with the canonical commutation relations of operators $b_{i}$
and $\overline{b}_{j}$, i.e.,
\begin{equation*}
\left[ b_{i},\overline{b}_{j}\right] _{\pm }=\delta _{ij}\text{; }\left[
b_{i},b_{j}\right] _{\pm }=0,
\end{equation*}%
the Hamiltonian $h_{\mathrm{Gen}}$ can act as a Hermitian one under the
biorthogonal basis \{$\left\vert n_{1},n_{2},...,n_{N}\right\rangle $, $%
\overline{\left\vert n_{1},n_{2},...,n_{N}\right\rangle }$\}. Here the
occupation number basis is defined as
\begin{eqnarray}
\left\vert n_{1},n_{2},...,n_{N}\right\rangle &\equiv &\prod_{l=1}^{N}\frac{%
(a_{l}^{\dag })^{n_{l}}}{\sqrt{n_{l}!}}\left\vert 0\right\rangle ,
\label{bio_basis} \\
\overline{\left\vert n_{1},n_{2},...,n_{N}\right\rangle } &\equiv
&\prod_{l=1}^{N}\frac{(\overline{b}_{l})^{n_{l}}}{\sqrt{n_{l}!}}\left\vert
0\right\rangle .  \notag
\end{eqnarray}

Thus, the solution of the Hamiltonian $h_{\mathrm{Gen}}$ can be obtained as
an ordinary Hermitian tight-binding model, which has the unchanged
topological structure. It is worthy pointing out that the Hamiltonian $H_{%
\mathrm{Gen}}$ with a complete set of biorthonormal eigenvectors is
pseudo-Hermitian \cite{AM}. Here we only consider the simplest case with $%
V_{j}=0$ and $\kappa =\sqrt{\mu \nu }$ (RT condition). One can see that the
Hamiltonian $h_{\mathrm{Gen}}$ reduces to an infinite uniform chain. Then a
plane wave of many particles still transmits completely through the
scattering center without any reflection. The reflectionless transmission
phenomenon can be well understood in this sense.

Moreover, such kind of transformation can be applied to other kinds of
finite one-dimensional system containing the UHA dimers to construct the
biorthogonal bases and reveal the exceptional point. To illustrate this
point, we will investigate two typical systems, ring and chain respectively.

\subsection{Finite chain system}

Now we consider a chain system which contains only a single UHA dimer. The
obtained conclusion can be extended to the case with multi-UHA dimer.
Without losing generality, we focus on the dimer with arbitrary $\left\{ \mu
,\nu \right\} $ rather than the RT dimer. In the following we consider the
case with fixed $\nu $\ but variable $\mu $. It allows us to simplify the
problem of the phase transition associated with the symmetry breaking.

For simplicity, the UHA dimer is embedded in the middle of chain with even
sites, which has the Hamiltonian%
\begin{eqnarray}
H_{\text{\textrm{Cs}}}=-\kappa \sum\limits_{j=1,j\neq N/2}^{N-1}\left(
a_{j}^{\dagger }a_{j+1}+\text{\textrm{H.c.}}\right) &&  \label{H_Cs} \\
-\mu a_{N/2}^{\dagger }a_{N/2+1}-\nu a_{N/2+1}^{\dagger }a_{N/2}. &&  \notag
\end{eqnarray}%
In the single-particle invariant subspace, the Bethe ansatz eigenwave
function can be expressed as%
\begin{equation}
f^{k}\left( j\right) =\left\{
\begin{array}{c}
Ae^{ikj}+Be^{-ikj},\text{ }j\in \left[ 1,N/2\right] \\
Ce^{ikj}+De^{-ikj},\text{ }j\in \left[ N/2+1,N\right]%
\end{array}%
\right.
\end{equation}%
By carrying out the similar procedure as above, we get the critical equation%
\textbf{\ }%
\begin{equation}
\Gamma _{\text{\textrm{Cs}}}\left( k\right) =\kappa ^{2}\sin ^{2}\left[
\left( \frac{N}{2}+1\right) k\right] -\mu \nu \sin ^{2}\left[ \left( \frac{N%
}{2}\right) k\right] =0,  \label{Gmma_Cs}
\end{equation}%
which determines the solution of the Hamiltonian. Here we concern the
critical behavior when the complex level appears. When the transition occurs
at energy level $k_{c}$,\ the following condition should be satisfied \cite%
{Liang Jin,L. Jin}%
\begin{equation}
\Gamma _{\text{\textrm{Cs}}}\left( k_{c}\right) =0\text{ and }\left. \frac{%
\partial \Gamma _{\text{\textrm{Cs}}}\left( k\right) }{\partial k}%
\right\vert _{k=k_{c}}=0.  \label{Critical_condition}
\end{equation}%
which yields the exceptional point $\mu =0$\ and $k_{c}=n\pi /\left(
N/2+1\right) $, $n\in \left[ 1,N/2\right] $. It indicates that the system
has fully real spectrum in the region $\mu \geqslant 0$, and all the $N/2$
pairs of energy levels coalesce at the exceptional point simultaneously. It
can also be demonstrated by the following critical behavior. Taking the
derivative with respect to the parameter $k$ on the Eq. (\ref{Gmma_Cs}), we
have%
\begin{equation}
\lim_{k\rightarrow k_{c}}\left( \frac{\partial \mu }{\partial k}\right) =0
\end{equation}%
which leads to%
\begin{equation}
\lim_{\mu \rightarrow 0}\left[ \frac{\partial E\left( k\right) }{\partial
\mu }\right] =-2\kappa \lim_{k\rightarrow k_{c}}\left[ \sin \left( k\right)
\frac{\partial k}{\partial \mu }\right] =\infty .  \label{repulsion}
\end{equation}%
It shows that the energy level exhibits the repulsive behavior when the
system tends to the exceptional point, which is in accordance with the
traditional non-Hermitian quantum mechanics.

This peculiar behavior cannot occur in a general Hermitian tight-binding
model. However, there should be a Hermitian model, referred to physical
counterpart, which has the identical real spectrum. It is interesting to
investigate what happens on the equivalent Hermitian model of the
non-Hermitian chain (\ref{H_Cs}) when the parameter $\mu $ approaches to the
exceptional point. To this end, one can apply the transformation
\begin{equation}
b_{j}\text{ }\left( \overline{b}_{j}\right) =\left\{
\begin{array}{cc}
\sqrt{\frac{\mu }{\nu }}a_{j}\text{ }\left( \sqrt{\frac{\nu }{\mu }}%
a_{j}^{\dagger }\right) & \left( N/2<j\right) \\
a_{j}\text{ }\left( a_{j}^{\dagger }\right) & \left( j\leqslant N/2\right)%
\end{array}%
\right. ,
\end{equation}%
on the Hamiltonian (\ref{H_Cs}) and yields%
\begin{eqnarray}
h_{\text{\textrm{Cs}}}=-\kappa \sum\limits_{j=1,j\neq N/2}^{N-1}\left(
\overline{b}_{j}b_{j+1}+\overline{b}_{j+1}b_{j}\right) &&  \label{hcs} \\
-\sqrt{\mu \upsilon }\left( \overline{b}_{N/2}b_{N/2+1}+\overline{b}%
_{N/2+1}b_{N/2}\right) . &&  \notag
\end{eqnarray}%
As pointed above, under the biorthogonal bases in the form of Eq. (\ref%
{bio_basis}), $h_{\text{\textrm{Cs}}}$\ is Hermitian, which can be employ to
investigate the critical behavior. We note that such a mapping $H_{\text{%
\textrm{Cs}}}\longrightarrow h_{\text{\textrm{Cs}}}$ is always available
except the exceptional point $\mu =0$.

For small $\left\vert \mu \right\vert \ll \kappa $, Hamiltonian $h_{\text{%
\textrm{Cs}}}$ represents two weakly coupled uniform identical chains of
size $N/2$. Since the Hermiticity of $h_{\text{\textrm{Cs}}}$\ under the
biorthogonal bases, one can employed the standard perturbation method to
investigate the property of the system in the vicinity of $\mu =0$. In the
single-particle subspace, the eigen wavefunction and energy of first-order
approximation can be expressed as%
\begin{eqnarray}
\left\vert \psi _{\pm }^{k}\right\rangle =\sqrt{\frac{2}{N+2}}\left\{
\sum_{j=1}^{N/2}\sin \left( kj\right) \right. && \\
+\left. \sqrt{\frac{\mu }{\upsilon }}\sum_{j=N/2+1}^{N}\sin \left[ k\left( j-%
\frac{N}{2}\right) \right] \right\} a_{j}^{\dagger }\left\vert \text{\textrm{%
Vac}}\right\rangle , &&  \notag \\
\varepsilon _{\pm }^{k}=-2\kappa \cos \left( k\right) \pm \frac{\sqrt{\mu
\upsilon }}{N+2}\sin k\sin \left( Nk/2\right) , &&  \notag
\end{eqnarray}%
where $k=2n\pi /\left( N+2\right) $, $n=1,...,N/2$. It shows that
eigenvalues are real in the region of $\mu >0$,\ and come in complex
conjugate pairs\ in the region of $\mu <0$. It accords with the conclusion
at which we have arrived. Furthermore, the eigenfunctions obey%
\begin{equation}
\mathcal{T}\left\vert \psi _{\pm }^{k}\right\rangle =\left\{
\begin{array}{cc}
\left\vert \psi _{\pm }^{k}\right\rangle , & \mu >0 \\
\left\vert \psi _{\mp }^{k}\right\rangle , & \mu <0%
\end{array}%
\right. ,
\end{equation}%
which shows that all eigenfunctions break the $\mathcal{T}$ symmetry, at
which the reality of the eigenvalues are lost. Then we conclude that it
possesses the same characteristic features with that of a $\mathcal{PT}$%
-pseudo-Hermitian system.

The results also provides an interpretation for the level repulsion in the
Hermitian counterpart. We find that for the Hermitian counterpart, the
non-analytic behavior at the exceptional point is not from the Jordan block,
which appears in the non-Hermitian model, but from the divergence of
derivatives of matrix elements. The similar situation also occurs in another
models \cite{ZXZ1,ZXZ2}, which may imply a general conclusion.

\subsection{Ring system}

Through the above discussion, we can make a conclusion that, for the case of
the above open boundary condition model, namely a chain system, which
contains one or more UHA dimers as shown in Figs. \ref{fig2}(b) and \ref%
{fig2}(c). The self-consistent transformation (\ref{bj}) always holds for $%
\mu \upsilon >0$, thus the reality of the spectrum can be achieved. However,
on a system with periodic boundary conditions, i.e., a ring system with the
RT dimers, the self-consistent transformation implies the unitarity of the
S-matrix of whole RT dimers, which are illustrated in Fig. \ref{fig2}(a). It
is presumable that the complex level should appear if the S-matrix of whole
RT dimers is not unitary as shown in Fig. \ref{fig2}(d). Moreover, such a RT
dimer can act as an amplifier, which induces a persistent amplification.
Thus one can infer that the existence of steady state is impossible. We
exemplify this point by a ring system with a single RT non-Hermitian dimer
as sketched in Fig. \ref{fig2}(d), and the Hamiltonian of the concerned
system can be written as
\begin{eqnarray}
H_{\mathrm{Rs}} &=&-\kappa \sum_{j=1}^{N-1}\left( a_{j}^{\dagger }a_{j+1}+%
\text{\textrm{H.c.}}\right)  \label{Hring} \\
&&-\mu a_{N}^{\dagger }a_{1}-\left( \kappa ^{2}/\mu \right) a_{1}^{\dagger
}a_{N}\text{.}  \notag
\end{eqnarray}%
In the single particle invariant subspace, the Bethe ansatz wave function
has the form%
\begin{equation}
\left\vert \psi _{k}\right\rangle =\sum_{j=1}^{N}f^{k}\left( j\right)
a_{j}^{\dagger }\left\vert 0\right\rangle \text{,}
\end{equation}%
where $f^{k}\left( j\right) $ is in the form of
\begin{equation}
f^{k}\left( j\right) =Ae^{ikj}+Be^{-ikj},j\in \left[ 1,N\right]
\end{equation}%
The explicit form of the Schr\"{o}dinger equation reads%
\begin{equation}
-\kappa f^{k}\left( j-1\right) -\kappa f^{k}\left( j+1\right) =Ef^{k}\left(
j\right) ,
\end{equation}%
for $j\in \left[ 2,N-1\right] $, and
\begin{eqnarray}
-\kappa f^{k}\left( N-1\right) -\mu f^{k}\left( 1\right) &=&Ef^{k}\left(
N\right) \text{,} \\
-\kappa f^{k}\left( 2\right) -\left( \kappa ^{2}/\mu \right) f^{k}\left(
N\right) &=&Ef^{k}\left( 1\right) \text{,}  \notag
\end{eqnarray}%
with the energy spectrum%
\begin{equation}
E=-\kappa \left( e^{ik}+e^{-ik}\right) .
\end{equation}%
The solution of $k$\ is determined by the critical equation
\begin{eqnarray}
\Gamma _{\mathrm{Rs}}\left( k\right) =\sin \left[ \left( N+1\right) k\right]
-\sin \left[ \left( N-1\right) k\right] && \\
-\left( \frac{\mu }{\kappa }+\frac{\kappa }{\mu }\right) \sin \left(
k\right) =0\text{,} &&  \notag
\end{eqnarray}%
which can be reduced as%
\begin{equation}
k=\frac{2m\pi }{N}+\theta \text{, }m\in \left[ 0,N-1\right]
\end{equation}%
with
\begin{equation}
\theta =\frac{1}{N}\left[ \arccos \left( \frac{\mu }{2\kappa }+\frac{\kappa
}{2\mu }\right) \right] .
\end{equation}%
One can see that $\theta $\ is always a complex number except the point $\mu
=\kappa $, which corresponds to the Hermitian uniform ring. We note that a
complex $\theta $\ can lead to a complex $k$, which results in a complex
energy level. It is worthy pointing out that, the complex energy level
becomes real\ in the case of Re$\left( k\right) =0$, which corresponds to a
bound state. Then we conclude that a uniform ring containing a RT dimer does
not have a fully real spectrum. This point can also be understood by the
operator transformation. One can not find a set of self-consistent
transformation as Eq. (\ref{bj}) to rewrite the original Hamiltonian in Eq. (%
\ref{Hring}) as any form of Hermitian tight-binding model. This conclusion
can be extended to the case of containing multi-RT dimers with a non-unitary
S-matrix. As we mentioned above, however, things are different for a chain.
We will demonstrate it in the two dimensional non-Hermitian system.


\begin{figure}[tbp]

\hskip -11.5mm
\includegraphics[bb=0 0 568 767,width=0.47\textwidth,
clip]{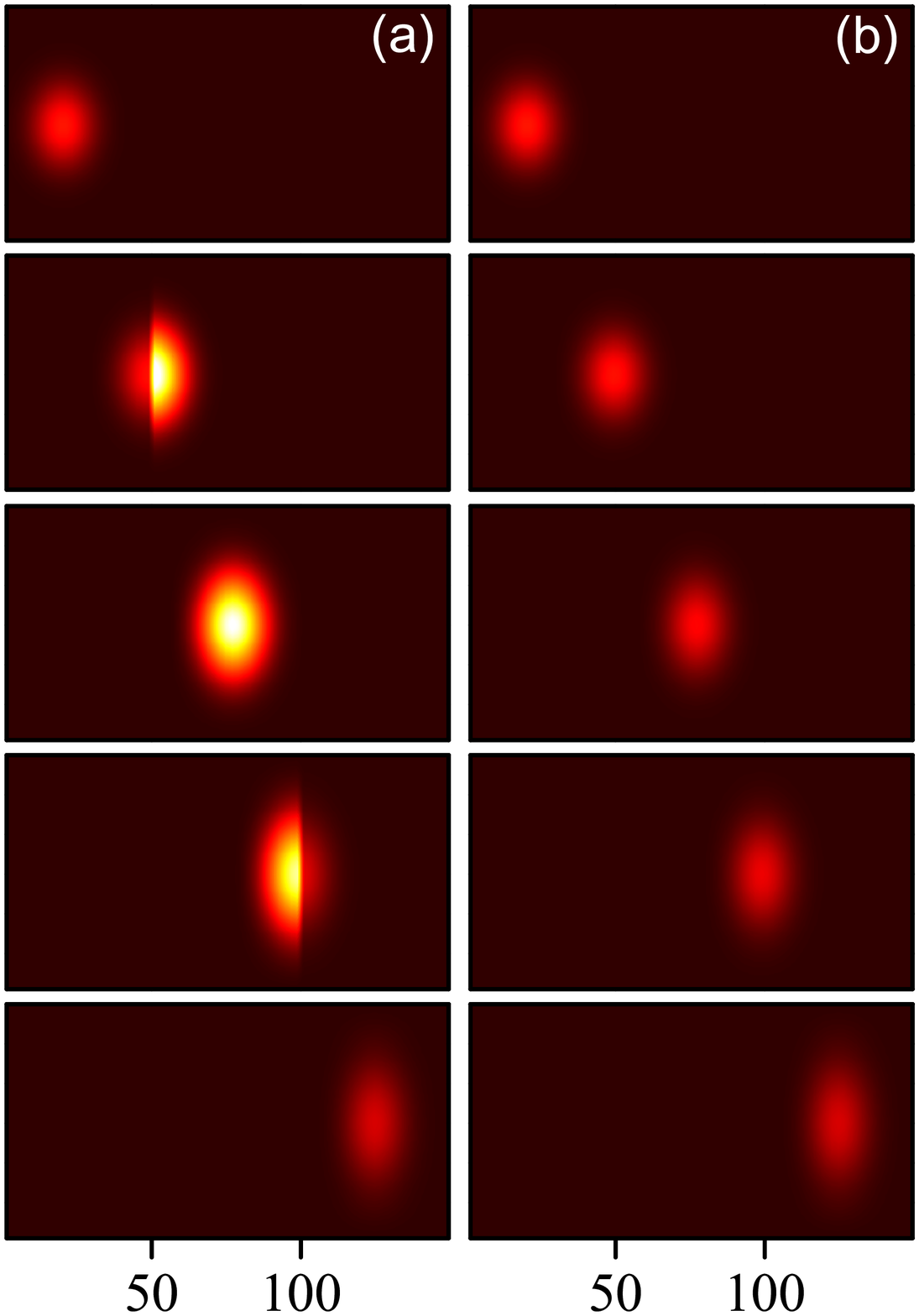}

\hskip -7.5mm
\includegraphics[ bb=19 197 583 547,width=0.47\textwidth,
clip]{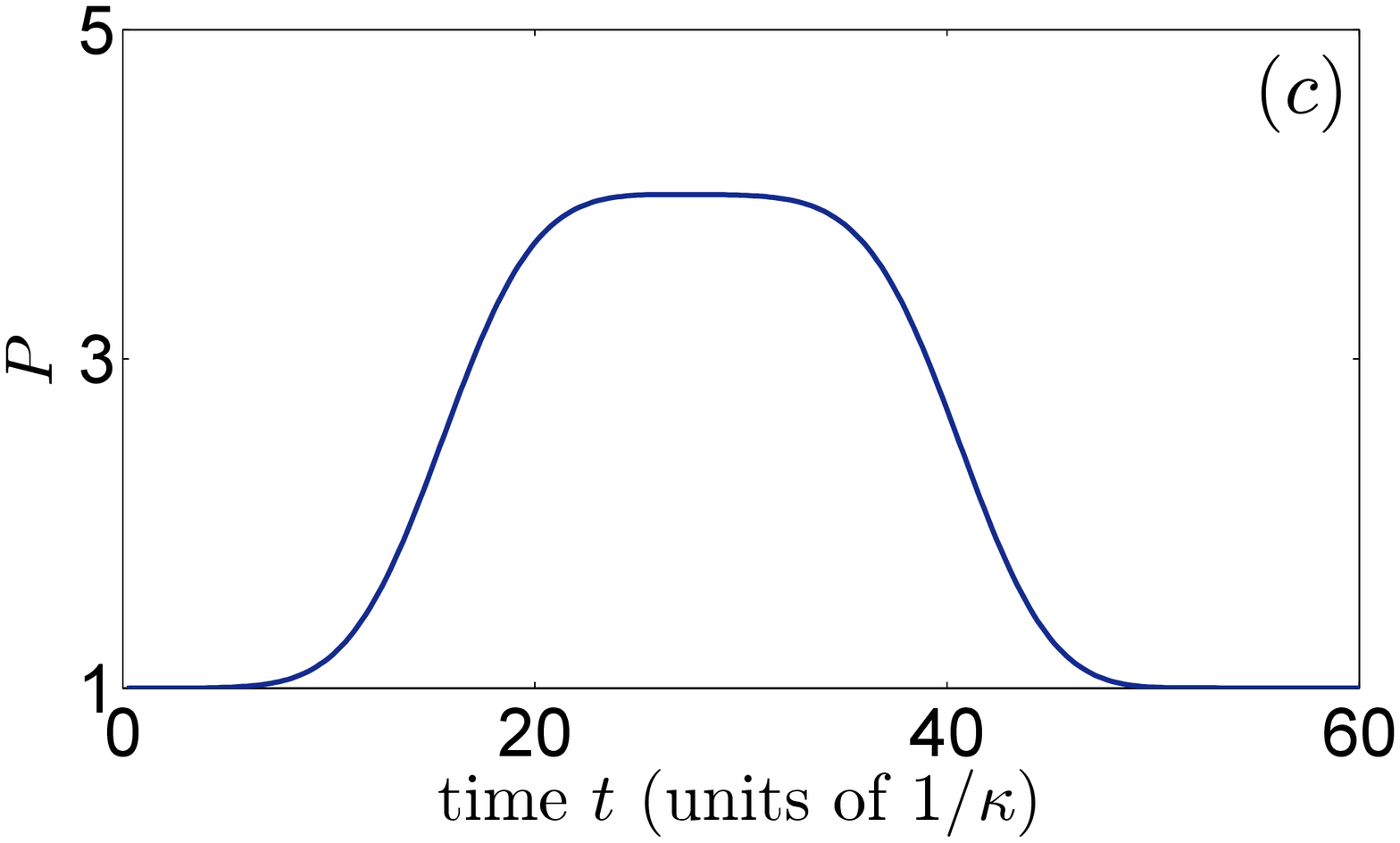}
\caption{(Color online) Propagation of an initial Gaussian-shaped wave
packet [snapshot of the site occupation probabilities $P_{l,j}\left(
t\right) =\left\vert \langle l,j|\protect\psi (t)\rangle \right\vert ^{2}$].
(a) in the two dimensional lattice with two non-Hermitian layers at $m$th
and $n$th shown in Fig. \protect\ref{fig3} (with parameter values given in
the text) and (b) in the defect-free lattice. It can be seen that the region
II does not scatter and absorb\ the incident wave packet, but instead it
propagates in such a way that it appears to the observers in regions I and
III as if there were no region II present. Then two layers can be regarded
as an invisible object. In (c) shows the behavior of the total occupation
probability $P(t)$ versus time in the two dimensional non-Hermitian lattice
of (a).} \label{fig4}
\end{figure}


\section{Wavepacket dynamics and invisibility}

\label{sec_dynamic}Based on the above analysis, one can extend the
conclusion to higher dimensional non-Hermitian systems. In this paper, we
focus on the application of such an extension, rather than providing a
universal proof. We consider a two-dimensional square lattice with $N$ sites
along the horizontal direction and $M$ sites along the vertical direction.
It contains two non-Hermitian layers at $m$th and $n$th columns $\left(
1<m<n<N\right) $, governed by the Hamiltonian

\begin{eqnarray}
&&H_{\mathrm{2D}}=H_{\mathrm{0}}  \notag \\
&&-\left( \mu -\kappa \right) \sum\limits_{i=1}^{M}\left( a_{i,m}^{\dagger
}a_{i,m+1}-\frac{\kappa }{\mu }a_{i,m+1}^{\dagger }a_{i,m}\right)  \notag \\
&&-\left( \mu -\kappa \right) \sum\limits_{i=1}^{M}\left( -\frac{\kappa }{%
\mu }a_{i,n}^{\dagger }a_{i,n+1}+a_{i,n+1}^{\dagger }a_{i,n}\right) ,
\label{HTD}
\end{eqnarray}%
where%
\begin{equation}
H_{\mathrm{0}}=-\kappa \sum\limits_{i=1,j=1}^{M-1,N}a_{i,j}^{\dagger
}a_{i+1,j}-\kappa \sum\limits_{i=1,j=1}^{M,N-1}a_{i,j}^{\dagger }a_{i,j+1}+%
\text{\textrm{H.c.}}
\end{equation}%
The two non-Hermitian layers are balanced in the sense that the S-matrix of
them is unitary. This can be shown from the following analysis. In Fig. \ref%
{fig3} the structure of the system is illustrated schematically. Two layers
separate the square into three regions. Performing the transformation of the
operators in these regions

\begin{equation}
b_{i,j}\text{ }\left( \overline{b}_{i,j}\right) =\left\{
\begin{array}{cc}
a_{i,j}\text{ }\left( a_{j}^{\dagger }\right)  & \left( j\geqslant
n+1\right)  \\
\frac{\mu }{\kappa }a_{i,j}\text{ }\left( \frac{\kappa }{\mu }%
a_{i,j}^{\dagger }\right)  & \left( m<j<n+1\right)  \\
a_{i,j}\text{ }\left( a_{i,j}^{\dagger }\right)  & \left( j\leqslant
m\right)
\end{array}%
\right. ,
\end{equation}%
the original Hamiltonian can be written as a simply form

\begin{eqnarray}
h_{\mathrm{TD}} &=&-\kappa \sum\limits_{i=1,j=1}^{M-1,N}\left( \overline{b}%
_{i,j}b_{i+1,j}+\overline{b}_{i+1,j}b_{i,j}\right) \\
&&-\kappa \sum\limits_{i=1,j=1}^{M,N-1}\left( \overline{b}_{i,j}b_{i,j+1}+%
\overline{b}_{i,j+1}b_{i,j}\right)  \notag
\end{eqnarray}

The canonical commutation relations \
\begin{equation}
\left[ b_{i,j},\overline{b}_{i^{\prime },j^{\prime }}\right] _{\pm }=\delta
_{ii^{\prime }}\delta _{jj^{\prime }}\text{; }\left[ b_{i},b_{j}\right]
_{\pm }=0,
\end{equation}%
ensure that the Hamiltonian $h_{\mathrm{TD}}$ can act as a defect-free
lattice and has the same spectrum and structure with the Hamiltonian $H_{%
\mathrm{0}}$. Furthermore, in regions I and III, we noticed that the
particle operators remain unchanged. Then the defect layers cannot be
detected (or invisible) for observers in regions I and III. We have checked
these predictions by direct numerical simulations of wave packet propagation
in Hermitian and non-Hermitian tight-binding systems of $H_{\mathrm{2D}}$
and $H_{\mathrm{0}}$. The initial state is a Gaussian-shaped wave packet in
the region I, which has the form%
\begin{eqnarray}
&&\left\vert \psi (0)\right\rangle =\frac{1}{\sqrt{\Omega _{0}}}%
\sum_{l,j}e^{i\left( k_{x}l+k_{y}j\right) }  \notag \\
&&\times e^{-\frac{\alpha ^{2}}{2}\left[ \left( l-N_{A}\right) ^{2}+\left(
j-N_{B}\right) ^{2}\right] }a_{l,j}^{\dagger }\left\vert 0\right\rangle ,
\end{eqnarray}%
It represents a wave packet located at $l$, $j$ site, with momentum $\left(
k_{x},k_{y}\right) $, where $\Omega _{0}=\sum_{i,j}\exp \{-\alpha ^{2}\left[
\left( i-N_{A}\right) ^{2}+\left( j-N_{B}\right) ^{2}\right] \}$\ is the
normalization factor and the half-width of the wave packet is $2\sqrt{\ln 2}%
/\alpha $.\textbf{\ }Fig. \ref{fig4}(a) shows that the propagation of an
initial Gaussian-shaped wave packet $\left\vert \psi (t)\right\rangle =\exp
\left( -iHt\right) \left\vert \psi (0)\right\rangle $ in a non-Hermitian
system of $H_{\mathrm{2D}}$ with the parameter $M=60$, $N=150$, $m=50$, $%
n=100$, $k_{x}=\pi /2$, $k_{y}=0$, $N_{A}=30$, $N_{B}=20$\ and $\alpha =0.1$%
\textbf{. }For comparison, Fig. \ref{fig4}(b) shows the propagation of the
same wave packet in the defect-free system of $H_{\mathrm{0}}$. It can be
seen that the total probability $P(t)=\sum_{l,j}\left\vert \langle l,j|\psi
(t)\rangle \right\vert ^{2}$ of the wave packet turns out to be amplified
(diminished) during interaction with the first (second) defect layer. It is
shown that the two non-Hermitian layers are balanced, which ensure the
probability $P(t)$ conserved in regions I and III. The defect layers in
addition of being reflectionless, are also invisible to the an outside
observer requires that the phase $\varphi \left( k\right) $ of the
transmission coefficient be flat, that is, $\left( d\varphi /dk\right) =0$
almost everywhere. If this condition is not satisfied, the spectral
components of a wave packet crossing the defect region of the partner
lattice would acquire the additional phase contribution $\varphi \left(
k\right) $, absent in the defect-free lattice, which would be responsible
for a different time of flight and for a different distortion of the wave
packet as compared to the same wave packet propagating in the ideal
defect-free lattice. The advance in the time of flight experienced by the
wave packet propagating in the lattice with defects can be readily
calculated by standard methods of phase or group-delay time analysis and
reads \cite{SLonghi}%
\begin{equation}
\tau _{\mathrm{g}}=\frac{1}{v_{\mathrm{g}}}\left( \frac{d\varphi \left(
k\right) }{dk}\right) _{k_{x}}=\frac{1}{2\kappa \sin \left( k_{x}\right) }%
\left( \frac{d\varphi \left( k\right) }{dk}\right) _{k_{x}}
\end{equation}%
where $k_{x}=\pi /2$ is the carrier wave number of the wave packet and $v_{%
\mathrm{g}}=2\kappa \sin \left( k_{x}\right) >0$ is its group velocity in
the $x$-direction. The characteristic of reflectionless and $k$-independent
transmission for the S-matrix of two layers leads to $\varphi \left(
k\right) =0$, which generates $\tau _{\mathrm{g}}=0$. Thus the wave packet
is fully transmitted with no appreciable delay and/or distortion. An outside
observer thus cannot distinguish whether the transmitted wave packet has
been propagated in a defect-free system $H_{\mathrm{0}}$ or in a
non-Hermitian system $H_{\mathrm{2D}}$ in regions I and III. It should be
noted that the total probability $P(t)$ is not conserved in the
non-Hermitian system $H_{\mathrm{2D}}$, as shown in Fig. \ref{fig4}(c). Such
the enhancement and diminishment of the probability $P(t)$, however, is not
visible to the outside observer.

\section{Summary}

\label{sec_summary}

In summary, we have investigated the non-Hermitian system which has $%
\mathcal{T}$ rather than $\mathcal{PT}$ symmetry, with the non-Hermiticity
arising from the UHA dimers. Just like a $\mathcal{PT}$ symmetric
Hamiltonian, the concerned system shares the same properties with that of a
pseudo-Hermitian Hamiltonian. The eigenvalues of the Hamiltonian are either
real or come in complex-conjugate pairs. We found that the corresponding
Hermitian counterpart can be obtained by a similarity transformation, which
is experimentally accessible system, with the unchanged topological
structure but modulated hopping amplitude. Comparing with other two types of
non-Hermitian clusters, complex on-site potential and non-hermitian dimer as
the scattering centers, the UHA dimer exhibits the peculiar feature that the
transmission coefficient of it is $k$-independent under the reflectionless
transmission condition. This fact indicates that a balanced combination of
the RT dimers can act as a Hermitian scattering center, leading to fully
transmission of an arbitrary wave packet with no appreciable delay and/or
distortion. The defects can appear fully invisible to an outside observer.
This paves an alternative way for the design of invisible cloaking devices.

\acknowledgments We acknowledge the support of National Basic Research
Program (973 Program) of China under Grant No. 2012CB921900.

\end{document}